\documentclass[osajnl,twocolumn, showpacs, superscriptaddress,10pt]{revtex4-1}   
\usepackage{amsmath,amssymb,graphicx}
\usepackage{epstopdf}

\begin{document}


\title{Sub-Doppler Cooling of Neutral Atoms in a Grating Magneto-Optical Trap}


\author{J. Lee}\email{neoscien@umd.edu}\author{J. A. Grover}\author{L. A. Orozco}\author{S. L. Rolston}

\affiliation{Joint Quantum Institute, Department of Physics, University of
Maryland and National Institute of Standards and Technology, College Park, Maryland 20742, USA}

\begin{abstract}
The grating magneto-optical trap (GMOT) requires only one beam and three planar diffraction gratings to form a cloud of cold atoms above the plane of the diffractors. Despite the complicated polarization arrangement, we demonstrate sub-Doppler cooling of $^{87}\mathrm{Rb}$ atoms to a temperature of 7.6(0.6) $\mu$K through a multi-stage, far-detuned GMOT in conjunction with optical molasses. A decomposition of the electric field into polarization components for this geometry does not yield a mapping onto standard sub-Doppler laser-cooling configurations. With numerical simulations, we find that the polarization composition of the GMOT optical field, which includes $\sigma$- and $\pi$-polarized light, does produce sub-Doppler temperatures.
\end{abstract}



\maketitle

\section{Introduction}
\label{sec:intro}

Laser cooling and trapping of neutral atoms in compact structures
has been studied for reliable, replicable, and portable setups in applications of quantum metrology and quantum information. Instead
of the standard, six-beam magneto-optical trap (MOT), a tetrahedral
MOT was realized with four beams \cite{Shimizu91}. A pyramid MOT
(PMOT) was demonstrated with a single beam incident on a corner
of four mirrors \cite{Lee96}, and integrated PMOT arrays were realized by etching pyramids into a silicon wafer \cite{Pollock09}.

Atom chips \cite{Reichel99} with atoms trapped above their mirror surfaces are easily integrated with other devices, such as fiber Fabry-Perot cavities \cite{Reichel07} or tapered-fiber-coupled microdisk cavities \cite{Painter06}. A single-beam tetrahedral MOT with three angled mirrors \cite{Vangeleyn09} or with three planar diffractors - a grating MOT (GMOT) \cite{Vangeleyn10} - combines the simplicity of a single beam with the ability to trap near a  surface.  A recently demonstrated GMOT with nanofabricated gratings achieves sub-Doppler cooling and higher atom number than earlier GMOT experiments \cite{Nshii13}. Compression of the diffracted GMOT beams along the beam axes increases their intensities, but the low diffraction efficiency of the correct handedness compensates this issue to produce the balanced forces needed for atom trapping.

Recent proposals in quantum information focus on hybrid quantum systems, given that no single architecture possesses the necessary attributes to have both fast quantum logic gates and a quantum memory with long coherence times \cite{Divincenzo00}.  One such hybrid quantum computing platform arises from coupling neutral atoms to superconducting circuits \cite{Verdu09, Hoffman11}.  This requires laser cooling and trapping the atoms in a dilution refrigerator.  The space and heating constraints of the refrigerator demand that any MOT placed inside of it must be compact and thermally separated from the lowest temperature region.  We are pursuing a combination of a GMOT with an optical nanofiber trap \cite{Vetsch10} in order to realize the atomic part of this hybrid system. The GMOT provides a compact laser cooling setup, and the nanofiber can transport optically trapped atoms from the GMOT to a few micrometers above the superconducting circuit. The nanofiber trap depth is on the order of $100\,\mu \textrm{K}$, and the trapping sites are in the collisionally blockaded regime \cite{Schlosser02}.  This necessitates sub-Doppler cooling and high atomic density for the efficient transfer of atoms from the GMOT to the nanofiber trap.

Using the single-beam GMOT with planar diffractors mounted on an ultra-high vacuum manipulator, we obtain sub-$10\,\mathrm{\mu K}$ temperatures through a multi-stage sub-Doppler cooling process. Our vacuum manipulator with micrometer translation stages helps to locate the optimal magnetic field value and beam intersection for the GMOT. The carefully pre-adjusted bias magnetic fields for each step of our sub-Doppler cooling process help to hold atomic clouds in the capture volume for
up to $10\,\mathrm{ms}$ during an optical molasses cooling stage.

Following the demonstration of optical molasses \cite{Chu85} and the discovery of sub-Doppler cooling temperatures \cite{Lett88}, the sub-Doppler cooling processes of lin$\perp$lin and $\sigma^{+}$-$\sigma^{-}$ 1D optical molasses were explained by the existence of  polarization gradients and the non-adiabatic response of moving atoms to the light fields \cite{Dalibard89, Ungar89}. Because of orthogonal polarizations, both cases have no intensity standing waves. For the lin$\perp$lin configuration, sub-Doppler cooling is explained by the ``Sisyphus effect," where atoms feel a m-state-dependent light shift potential and lose energy by climbing a potential hill and then optically pumping to a different m-state, returning to the bottom of the hill. Sisyphus cooling requires a ground state angular momentum, $J_g\geq 1/2$.  For the $\sigma^{+}$-$\sigma^{-}$ polarization configuration, sub-Doppler cooling arises from velocity-selective Raman transitions that efficiently transfer population from one m-state to another.  Because it relies on Raman transitions using photons from each beam, it  requires $J_g\geq 1$.  Another version of Sisyphus cooling exists for a $\sigma$-standing wave in a weak transverse magnetic field,  called  magnetic-field-induced laser cooling \cite{Sheehy90}. The state- and spatially-dependent light shift of a standing wave, optical pumping at the anti-node, and Zeeman substates mixing in the absence of any light produce Sisyphus cooling without a polarization gradient.

The GMOT optical configuration does not obviously map onto any of the  sub-Doppler cooling mechanisms described above. The light reflected from the gratings nominally retains its handedness, but the propagation axis changes. The result is a combination of linear and circular polarizations, with polarization and intensity gradients. A full theoretical treatment of this 3D configuration is beyond the scope of this paper.  We present a simplified 1D model with polarizations and intensities appropriate for translations along symmetry axes of the optical field.  We find that the existence of $\pi$-polarized light in addition to $\sigma$-polarized light does provide a mechanism for sub-Doppler cooling. Numerical calculations determine the existence of a sub-Doppler cooling feature near zero velocity for the GMOT with the well-defined polarization states and the effective wavevectors of the selected 1D optical lattices.

We discuss the experimental setup of our GMOT configuration in Sec.~\ref{sec:setup}; Sec.~\ref{sec:temp} presents the temperature of the sub-Doppler laser-cooled atoms as a function of the detuning of the cooling beam; the simulation of the sub-Doppler cooling process of the GMOT is in Sec.~\ref{sec:theory}; Sec.~\ref{sec:number} presents atom number and atomic density with sub-Doppler cooling,  as well as atom number without sub-Doppler cooling, as a function of the magnetic field gradient; and Sec.~\ref{sec:conc} presents the conclusions.

\section{Experimental Setup}
\label{sec:setup}

\begin{figure}[h!]
\centering\includegraphics[scale=0.4]{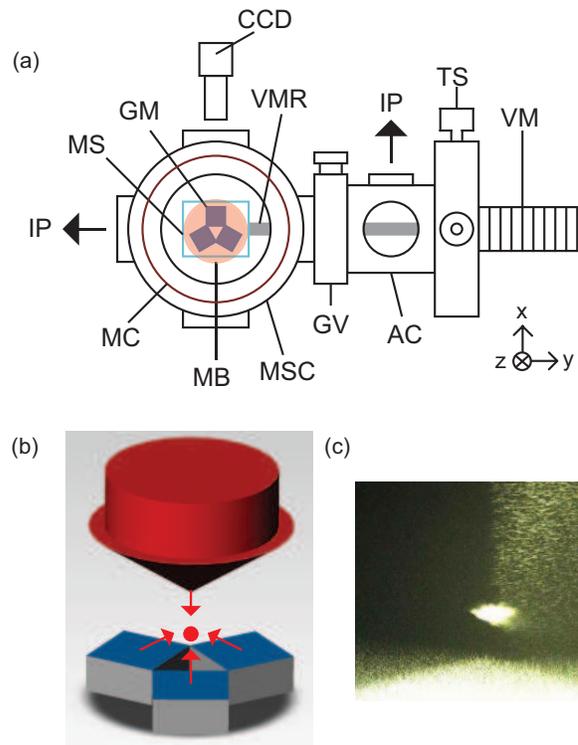}
\caption{(color online) (a) Experimental setup. \textbf{AC}: Antechamber; \textbf{CCD}: CCD
camera; \textbf{GM}: Grating Mirrors; \textbf{GV}: Gate Valve; \textbf{IP}: Ion Pump; \textbf{MB}: MOT Beam; \textbf{MC}: MOT Coils; \textbf{MS}: Microscope Slide; \textbf{MSC}: Main Science Chamber; \textbf{TS}: 2-D Translation Stage; \textbf{VM}: Vacuum Manipulator; \textbf{VMR}: Vacuum Manipulator Rod. (b) Single-beam GMOT configuration. (c) GMOT atom image with CCD camera.}
\label{fig1}
\end{figure}

Our vacuum chamber is composed of the main science chamber (MSC),
the antechamber (AC), and a vacuum manipulator (VM). A gate value (GV) connects the main chamber
and antechamber, and the end of the rod of the vacuum manipulator
(VMR) holds the grating mirrors, as shown in Fig.~\ref{fig1} (a). This vacuum system is convenient
for testing various setups without breaking the vacuum of the main chamber, reducing possible contamination of the science chamber. Closing the gate valve and opening the antechamber, we install
a setup and pump down the antechamber with a turbo pump. We reopen
the gate valve and transfer the setup into the main chamber using
the vacuum manipulator. Ion pumps (Duniway DSD-050-5125-M, $50\,\mathrm{L\cdot s^{-1}}$)
are attached to the main chamber and the antechamber.  We trap from the vapor coming from a Rb dispenser (SAES Getters, FT type) that thermalizes with the walls of the chamber after Rb coats the surface. The vacuum pressure during these measurements is about $10^{-9}\,\mathrm{mbar}$,
and the $1/e$ atomic loading time is about $2-3\,\mathrm{sec}$.

The vacuum manipulator (VG Scienta Transax) is composed of a 1D motorized vacuum manipulator
(VM, $\mathbf{z}$-axis) and 2-D manual translation stages (TS, $\mathbf{xy}$-axes), as shown in
Fig. \ref{fig1} (a). The vacuum manipulator holds the gratings and
can precisely adjust the position of the gratings ($450\,\mathrm{mm}$ total z translation with $5\,\mathrm{\mu m}$ resolution and $25\,\mathrm{mm}$ vectorial xy translation with $5\,\mathrm{\mu m}$
resolution) to find the optimal magnetic field value and beam overlap to reach the lowest temperature. Fig.~\ref{fig1} (b) shows the arrangement of the three commercial gratings that we use ($12.7\,\mathrm{mm}\times12.7\,\mathrm{mm}\times6\,\mathrm{mm}$,
Edmund  NT43-752, $1200\,\mathrm{grooves/mm}$). The incoming beam overlapping with the first-order reflections from the three gratings generates a capture volume of $\sim100\,\mathrm{mm^{3}}$ with a single beam that is spatially filtered with a single-mode optical fiber. Three gratings are glued
on microscope slides (MS) with UV epoxy (EPO-TEK OG116-31), and the slides are attached to the support rod of the manipulator. The microscope slides are arranged such that there is a gap in the middle of the gratings to prevent reflections that cause force imbalances in the MOT.  We tune our lasers to the $D_2$ line (780 nm) of $^{87}\mathrm{Rb}$, which has a natural linewidth of $\Gamma\,(=2\pi\times 6$ MHz). The cooling beam ($I=2.5\,\mathrm{mW/cm^{2}}$) is locked to the $F=2$
to $F'=3$ transition, and the repumper beam
($I=0.2\,\mathrm{mW/cm^{2}}$) is locked to the $F=1$ to $F'=2$ transition. Both beams are sent through the same fiber.  The polarization of the single cooling beam is circular, but the polarization of the first-order diffraction changes. The first-order diffraction efficiency for the MOT beam is $30(5)\,\%$, leading to balanced optical molasses \cite{Vangeleyn10}. A GMOT requires a large and high quality beam. 
We expand the beam directly out of a single-mode optical fiber to a diameter to $3.6\,\mathrm{cm}$, and this beam is collimated with a shearing interferometer. We then finely align the beam with a tiltable mount and 3D translation stage to optimize the GMOT.

\section{Temperature measurement}
\label{sec:temp}

\begin{figure}[h!]
\centering\includegraphics[scale=0.4]{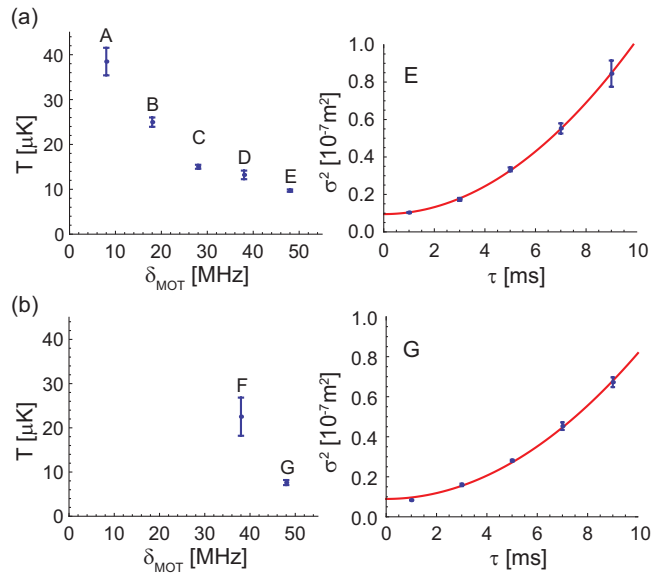}
\caption{(color online) (a) Temperature versus the detuning of the cooling beam for a single-stage, far-detuned GMOT (see Table~\ref{tab1}); $\sqrt{2}\sigma$ is the $1/e$ radius of atomic cloud, and we fit $1-9\,\mathrm{ms}$ time-of-flight data to $\sigma^{2}=\sigma_{0}^{2}+\sigma_{v}^{2}t^{2}$ (right). We estimate the atomic temperature $T\,(=m_{Rb}\sigma_{v}^{2}/k_{B})$ from the fits (left). (b) Temperature versus the detuning of the cooling beam for a multi-stage, far-detuned GMOT with no molasses stage (F) and with a molasses stage (G) (see Table~\ref{tab2}).}
\label{fig2}
\end{figure}

Measuring the mean square radius of the two dimensional cloud image versus expansion time allows us to extract the atomic temperature ($T=m_{Rb}\sigma_{v}^{2}/k_{B}$) from fits of $\sigma^{2}=\sigma_{0}^{2}+\sigma_{v}^{2}t^{2}$. Fig.~\ref{fig2} displays the results of this measurement for different experimental conditions. Fig.~\ref{fig2} (a, left) presents atomic cloud temperatures as measured after cooling for $50\,\mathrm{ms}$ in a single-stage, far-detuned GMOT (see Table~\ref{tab1}). The lowest observed temperature for this procedure is $9.7\,(0.3)\mathrm{\mu K}$ and occurs at a detuning of $8.2\,\Gamma$ (an example of the fit for this detuning is given in Fig.~\ref{fig2} (a, right)).

Employing a molasses cooling stage after a multi-stage, far-detuned
GMOT requires the adjustment of the GMOT position as the magnetic
field gradient decreases to zero (Fig. \ref{fig2} (b) and Table~\ref{tab2}). We optimize our bias magnetic field for each far-detuned GMOT stage such that the
laser-cooled atoms remain in the capture volume for up to $10\,\mathrm{ms}$
after turning off the magnetic field. Table~\ref{tab2} summarizes the steps in this process. In the multi-stage, far-detuned GMOT procedure without molasses, we measure an atomic temperature of $22.5(4)\,\mathrm{\mu K}$ (Fig. \ref{fig2} (b, left), F). This temperature can be explained by the final detuning of $6.5\,\Gamma$ being closer to the resonance than that of the single-stage, far-detuned MOT. The atomic temperature after the multi-stage, far-detuned GMOT and a $1\,\mathrm{ms}$ molasses stage (at a detuning of $8.2\,\Gamma$) is $7.6(0.6)\,\mathrm{\mu K}$ (Fig. \ref{fig2} (b, left), G), which is colder than the single-stage, far-detuned GMOT without molasses described above. 

\begin{table}
	\begin{tabular}{|c | c | c | c|}\hline
		Cooling time $\tau$ [ms] & $\delta_{MOT}/\Gamma$ & $dB/dz\,\mathrm{\left[G\cdot cm^{-1}\right]}$ & $I/I_{sat}$\\ \hline\hline
		$\tau_{MOT}$ & 1.5 & 10.8 & 1.55\\ \hline
		50 & $1.5\;\mathrm{to}\;8.2$ & 10.8 & 1.16\\ \hline
	\end{tabular}
\caption{Single-stage, $50\,\mathrm{ms}$ far-detuned MOT parameters when scanning the detuning of a single MOT beam (time flows downwards in the table). This table corresponds to Fig. \ref{fig2} (a); in this paper, the cooling process with several stages is represented
by the cooling stage time ($\tau$), the detuning of the cooling beam ($\delta_{MOT}$,
red-detuned from the cooling transition), magnetic field gradient
($dB/dz$), and the relative intensity of the single incident cooling beam ($I/I_{sat}=2\Omega^{2}/\Gamma^{2}$).} 
\label{tab1}
\end{table}

\begin{table}[h!]
	\begin{tabular}{|c | c | c | c|}\hline
		Cooling time $\tau$ [ms] & $\delta_{MOT}/\Gamma$ & $dB/dz\,\mathrm{\left[G\cdot cm^{-1}\right]}$ & $I/I_{sat}$\\ \hline\hline
		$\tau_{MOT}$ & 1.5 & 10.8 & 1.61\\ \hline
		30 & 3.2 & 10.8 & 1.40\\ \hline
		15 & 4.9 & 6.6 & 1.20\\ \hline
		15 & 6.5 & 4.5 & 1.20\\ \hline
		1 & 8.2 & 0 & 1.20\\ \hline
	\end{tabular}
\caption{Multi-stage, $60\,\mathrm{ms}$ far-detuned MOT and $1\,\mathrm{ms}$ optical molasses parameters, with time flowing downwards in the table. This table corresponds to Fig. \ref{fig2} (b) G; the same multi-stage far-detuned MOT without $1\,\mathrm{ms}$ optical molasses corresponds to Fig. \ref{fig2} (b) F.} 
\label{tab2}
\end{table}

\section{Theory}
\label{sec:theory}

Reference~\cite{Vangeleyn09} describes the requirements for magneto-optical trapping in a GMOT. They consist of finding a configuration where the optical forces sum to zero. We are interested in understanding sub-Doppler cooling in the polarization configuration present in the GMOT, as it is neither Sisyphus (lin$\perp$lin) polarization gradient cooling, nor $\sigma^+-\sigma^-$ orientational cooling. To simplify the theory, we will assume the cold atoms are close enough to the center of the quadrupole field so that we can neglect any Zeeman contribution to the laser detuning. We will consider only 1D laser cooling.

\begin{figure}
\centering\includegraphics[scale=0.4]{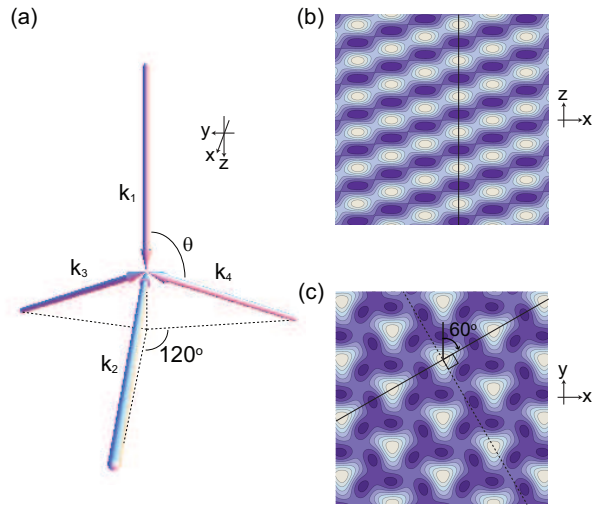}
\caption{(color online) (a) Tetrahedral MOT configuration; $|\cos\,\theta|=1/3$ and $\sum_{i=1}^{4} \mathbf{k}_{i} I_{i} = 0$. (b) Optical lattices in the $\mathbf{xz}$ plane (solid line: 1D optical lattice along the $\mathbf{z}$-axis of the $\mathbf{xz}$ plane). (c) Optical lattices in the $\mathbf{xy}$ plane (solid line: 1D optical lattice along the diagonal axis with an angle of $60^o$ relative to the $\mathbf{y}$-axis, dashed line: 1D optical lattice along the diagonal axis with an angle of $150^o$ relative to the $\mathbf{y}$-axis). }
\label{fig3}
\end{figure}

There is a stable polarization configuration (relative phases between beams displace the polarization configuration but do not change its morphology) in a four-beam configuration such as the GMOT. The spatial periodicity of the underlying lattice is determined by the geometry of four beams, with a primitive unit cell ($\mathbf{k}_{i}-\mathbf{k}_{j}$) of the reciprocal lattice \cite{Petsas94}, where $\mathbf{k}_{i}$ is the wavevector of the 3D beams (see Fig. \ref{fig3}).

The polarization pattern of the GMOT configuration is complicated because of the existence of both linear and circularly polarized light. For a chosen quantization axis along the vertical axis ($\mathbf{z}$), when the $\sigma$-polarized vertical beam reflects off the diffraction gratings, the handedness of the polarization (seen from the opposite direction of propagating beams with $\mathbf{k}$-vectors) is maintained (at the $\approx 90\,\%$ level), but in terms of the quantization axis, the reflected beams will have $\sigma^+,\ \sigma^- , \  {\rm and} \  \pi$ components. The exact composition can be calculated by a suitable transformation matrix that connects the axes through a rotation. In Fig \ref{fig3} (a-b), a crystal axis parallel to the vertical GMOT beam with its wavevector $\mathbf{k}_{1}$ has an angle of $109.5^o$ from three other GMOT beams with their wavevectors of $\mathbf{k}_2$, $\textbf{k}_3$, and $\textbf{k}_4$. For the quantization axis $\mathbf{q}_z$=(0,0,1), the polarization states of the vertical GMOT beam and the three GMOT beams projected along the vertical axis correspond to $100\,\%, 0\,\%, \,  {\rm and} \, 0\,\%$; $44.4\,\%, 11.1\,\%, \,  {\rm and} \, 44.4\,\%$ of $\sigma^+,\  \sigma^- , \ {\rm and}\  \pi$ respectively. In the horizontal ($\mathbf{xy}$) plane, the line at $60^o$ and its perpendicular at $150^o$ from the $\mathbf{y}$-axis define crystal axes in the system (Fig \ref{fig3} (c)). For the quantization axis $\mathbf{q}_{xy}$ = ($\sqrt{3}$/2,1/2,0), the polarization states of the three beams projected to the horizontal plane with $\mathbf{k}_2$, $\textbf{k}_3$, and $\textbf{k}_4$ correspond to $82.5\,\%, 0.8\,\%, \,  {\rm and} \, 16.7\,\%$; $25\,\%, 25\,\%, \,  {\rm and} \, 50\,\%$; $0.8\,\%, 82.5\,\%, \,  {\rm and} \, 16.7\,\%$ of  $\sigma^+,\  \sigma^- , \ {\rm and}\  \pi$ respectively. A simple retroflection of a circularly polarized beam would result in a standing wave without any polarization gradients and no sub-Doppler cooling. Additional polarization components due to the reflection angles are critical for a sub-Doppler cooling mechanism. 

\begin{figure}
\centering\includegraphics[scale=0.4]{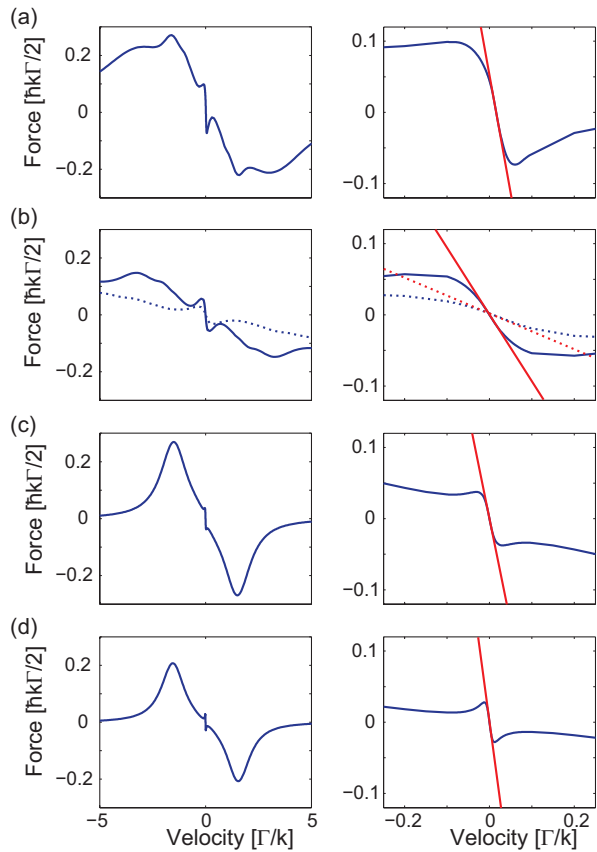} 
\caption{(color online) Calculated force on atoms as a function of atom velocity for different axes and polarization configurations of the GMOT. (a) Vertical axis of the GMOT, the $\mathbf{z}$-axis of the $\mathbf{xz}$ plane (Fig. \ref{fig3} (b)). (b) Horizontal axes of the GMOT, the diagonal axes of the $\mathbf{xy}$ plane. solid line: the axis at an angle of $60^o$ relative to the $\mathbf{y}$-axis, dashed line: the axis at an angle of $150^o$ relative to the $\mathbf{y}$-axis (see Fig. \ref{fig3} (c)). (c) $\sigma^{+}$-$\sigma^{-}$ orientational cooling. (d) lin-$\perp$-lin Sisyphus cooling where $I/I_{sat}$=1.2 and $\delta$=-1.5$\,\Gamma$. The right column shows a zoom of the region where the slope is largest around zero velocity.}
\label{fig4}
\end{figure}

We numerically calculate the force on the atoms versus atom velocity  along  the $\mathbf{z}$-axis of the $\mathbf{xz}$ plane (Fig. \ref{fig3} (b)) and the diagonal axes of the $\mathbf{xy}$ plane (Fig. \ref{fig3} (c)).  In the simulation, we include the multi-level structure of a $^{87}$Rb atom, such as the transitions from the $F=2$ Zeeman sub-states to the $F^{'}=3$ Zeeman sub-states. The steady state solution of the master equation, $\frac{d\hat{\rho}}{dt} = -\frac{i}{\hbar}[\hat{H},\hat{\rho}] + \Gamma_{\hat{\rho}}$, is solved by the matrix continued fraction method \cite{Minogin79}. Treating the beams as classical optical fields, the raising ($\hat{A}_{+}^{\dagger}$, $\hat{A}_{-}^{\dagger}$, $\hat{A}_{0}^{\dagger}$) and the lowering ($\hat{A}_{+}$, $\hat{A}_{-}$, $\hat{A}_{0}$) operators correspond to the optical pumping and spontaneous emission of the transitions of $\sigma^{+}$-, $\sigma^{-}$-, and $\pi$-polarized lights, respectively, and the Clebsch-Gordon coefficients for those operators define the transition strength between each hyperfine ground and excited state. The atom-light interaction Hamiltonian is $\hat{H}_{int} = -\frac{1}{2}(\Omega_{+}(\mathbf{r}) \hat{A}_{+} + \Omega_{-}(\mathbf{r}) \hat{A}_{-} + \Omega_{0}(\mathbf{r}) \hat{A}_{0}) + h. c.$, where $\Omega_{+}$, $\Omega_{-}$, and $\Omega_{0}$  are Rabi frequencies for $\sigma^{+}$-, $\sigma^{-}$-, and $\pi$-polarized lights, respectively. The force operator is $\hat{F} = -\nabla(\hat{H}_{int}$). After using the master equation to calculate the expectation value of $\hat{F}$ as a function of atomic velocity, we observe a sub-Doppler cooling signature (a steep slope of force vs. velocity) at low atom velocities for both crystal axes (Fig. \ref{fig4} (a) and (b)).

We calculate the force for different combinations of polarization in order to understand its role. If we have imbalanced $\sigma$-polarizations with no $\pi$-component, the narrow feature is present, but the point of zero force may not be contained within the feature, preventing sub-Doppler temperatures. This can be understood in the following way: there is orientational cooling for the part of the $\sigma^+$ component that balances the $\sigma^-$ component present, and then the force versus velocity curve is displaced vertically by the remaining unbalanced $\sigma^+$ component. When  there is $\pi$ polarization present, we recover a force versus velocity curve that should produce good sub-Doppler cooling with a narrow velocity feature centered on the zero-force point. This arises from coherent, two-photon, velocity-selective resonances between ground-state sublevels, coherent two-photon Raman transitions of $\sigma^+$-$\pi$ and $\pi$-$\sigma^-$ that become resonant when the energy difference between two sublevels is equal to the sum of opposite Doppler shifts of the two laser beams. The simulations show that the narrow velocity feature shifts horizontally away from zero velocity when a longitudinal magnetic field is present, similar to traditional $\sigma^+$-$\sigma^-$ orientational cooling \cite{Walhout92}. The horizontal shift of the force versus velocity curve is also accompanied by a vertical displacement as the magnetic field increases and negates the sub-Doppler cooling at higher magnetic fields.

 Figure 4 shows results of our model for different axes and polarization configurations present in the GMOT. The left column of the figure shows the broad features, while the right is a zoom on the region around zero. The atomic temperature $T$\,($=D_p/k_B\alpha$) is determined by the momentum diffusion coefficient $D_p$, related to heating and the momentum friction coefficient $\alpha$, related to cooling. If the spacing of the 1D optical lattice becomes more dense for a constant $D_p$, $\alpha$ increases because of the more frequent cooling events (note that our simulation does not calculate $D_p$, so we are unable to calculate actual temperatures). Assuming an isotropic diffusion constant, we expect the atomic temperature in the vertical direction, $T_{z}$, to be colder than the temperature in the horizontal direction, $T_{xy}$, based on the steeper slope of the force curve near zero velocity (see Figs.~ \ref{fig4} (a-b), where $I/I_{sat}$=1.2 and $\delta$=-1.5$\,\Gamma$). 

 In the experimental run with a multi-stage, far-detuned GMOT and a 1ms optical molasses (Fig. \ref{fig2} (b) G), $T_{z}$ is 1.5(0.25) times lower than $T_{xy}$; a recent GMOT experiment observes similar anisotropic sub-Doppler cooling \cite{Nshii13}. Given our 1D simplification, this can be considered a qualitative agreement. As a reference to compare the force vs. velocity features, we simulated the sub-Doppler cooling process of $\sigma^+$-$\sigma^-$ orientational cooling and lin$\perp$lin polarization gradient cooling (Fig. \ref{fig4} (c-d)). The slopes of vertical direction GMOT and $\sigma^+$-$\sigma^-$ cooling are similar (Fig. \ref{fig4} (a) and (c)). In addition, the amplitude of the Doppler cooling feature to capture atoms along the horizontal direction is lower than in the vertical direction because the intensities of the three GMOT beams along the horizontal axis are reduced compared to those along the vertical axis. Experimentally, we also observe an atom cloud squeezed along the vertical direction (Fig. \ref{fig1} (c)). If we assume comparable diffusion constants between the GMOT and traditional sub-Doppler mechanisms, our expectations and observations are similar.

\begin{figure}[h!]
\centering\includegraphics[scale=0.4]{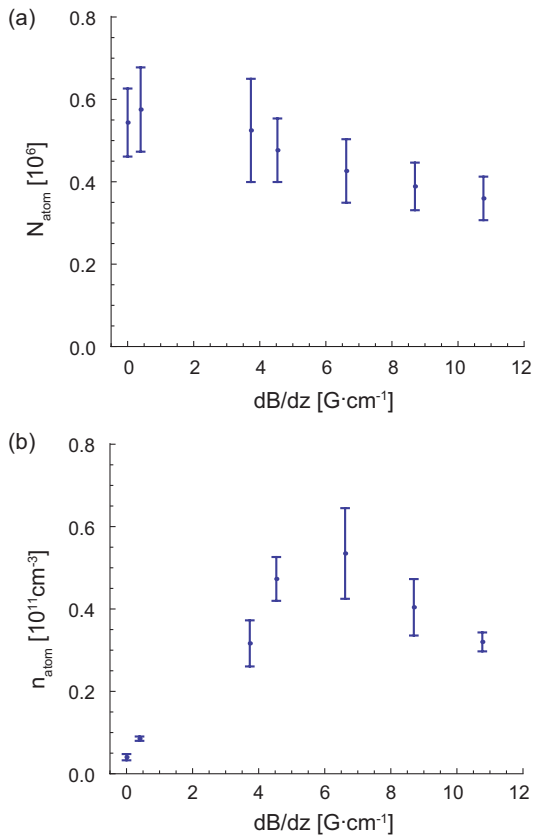}
\caption{(color online) Atom number ($N_{atom}$) and atomic peak density ($n_{atom}$) as a function of the magnetic field gradient ($dB/dz$).  Each data point has the same initial MOT atom number extracted simultaneously from a series of trials, and we vary $dB/dz$ during the far-detuned MOT process.}
\label{fig5}
\end{figure}

\section{Atom number and atom density}
\label{sec:number}

The success of loading cold atoms into the small (order $\lambda$) wells around the nanofiber requires many cold atoms at high density. We next study atom number ($N_{atom}$) and atomic peak density ($n_{atom}$) as a function of magnetic field gradient (Fig. \ref{fig5}) after cooling the atoms for $50\,\mathrm{ms}$ in a far-detuned GMOT. The experimental parameters are: ($\tau$ [ms], $\delta_{MOT}/\Gamma$, $dB/dz\,\mathrm{\left[G\cdot cm^{-1}\right]}$, $I/I_{sat}$) = ($\tau_{MOT}$, $1.5$, $10.8$, $1.29$) $\rightarrow$ ($50$, $3.9$, $0.4\;\mathrm{to}\;10.8$,
 $0.96$). Then, as $dB/dz$ increases, $N_{atom}$ decreases,
as seen in Refs.~\cite{Dalibard95,Walker90}. In addition, $n_{atom}$
also increases linearly as a function of $dB/dz$, but at
a certain peak density, the linear scaling does
not work any more because the light pressure from reradiated photons
limits the atomic density \cite{Dalibard95,Walker90,Lindquist92}. Reabsorption
of scattered photons within the trapped cloud becomes important above
$10^{11} \mathrm{atoms}\cdot\mathrm{cm}^{-3}$. In this regime, $n_{atom}$, which is nearly independent of $N_{atom}$, cannot be simply modeled due to the effective repulsive force between atoms. The atomic density decreases above the peak density because the multiple scattering of photons prevents further compression of the atomic cloud. Multiple scattering results in the heating of
atoms because of increased momentum diffusion and reduced friction
even with the restoring and friction forces of sub-Doppler cooling
\cite{Dalibard95}.

\begin{figure}
\centering\includegraphics[scale=0.4]{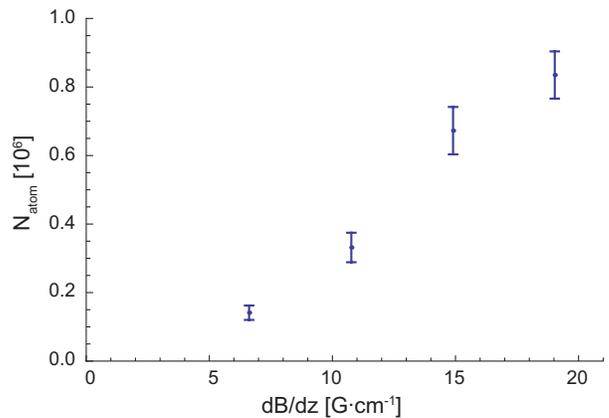}
\caption{(color online) Atom number in a GMOT (without sub-Doppler cooling) as a function of the magnetic field gradient.}
\label{fig6}
\end{figure}

A GMOT with no sub-Doppler cooling captures more atoms from the background atomic vapor as we increase the magnetic
field gradient (see Fig. \ref{fig6}). The capture velocity of the
GMOT increases when the magnetic fields in the GMOT shift the energy levels of atoms entering the trap from all directions. This is a mechanism
similar to that of a Zeeman slower, which has a spatially-varying
magnetic field to tune the atoms back into resonance as they decelerate and their Doppler shift changes. The parameters of the experiments are: ($\tau$ [ms], $\delta_{MOT}/\Gamma$, $dB/dz\,\mathrm{\left[G\cdot cm^{-1}\right]}$, $I/I_{sat}$) = ($\tau_{MOT}$, $1.5$, $6.6\;\mathrm{to}\;19$, $1.35$). The total number of atoms is smaller by two orders of magnitude than in typical MOTs.

\section{Conclusions}
\label{sec:conc}

We realize a single-beam tetrahedral GMOT with planar diffractors
that achieves sub-Doppler laser-cooling below $10\,\mathrm{\mu K}$
and traps $5\times10^{5}$ atoms. We also confirm the sub-Doppler cooling process in a 3D tetrahedral optical molasses following a multi-stage, far-detuned
GMOT required to reach low temperatures. We analyze the sub-Doppler cooling in a GMOT by projecting onto 1D axes and calculating the force as a function of atom velocity, recovering narrow cooling features. The sub-Doppler cooling arises from Raman processes between $\pi -$ and $\sigma -$polarized components of the optical field in this beam configuration. We study atom number and atomic peak density as a function of trap parameters and find densities large enough that rescattering becomes significant.  Based on our studies, the single-beam, tetrahedral MOT with planar diffractors has the potential to be a compact source of cold atoms that integrates well with surfaces, such as atom chips,  fiber-gap cavities \cite{Reichel07}, or nanofiber optical traps \cite{Vetsch10} for applications in quantum information science.

\medskip{}

Research supported by NSF through the PFC@JQI and the ARO Atomtronics MURI. We thank
Erling Riis for useful discussions.

\end{document}